\documentclass[12pt]{article}
\usepackage{amsmath}
\usepackage[dvips]{graphicx}
\usepackage{epsfig}
\usepackage{amsmath}
\usepackage{amssymb}
\usepackage{graphics,epstopdf}
\usepackage{amsfonts}
\usepackage{amsthm}
\usepackage[usenames]{color}

\begin{document}

\title{On the Mannheim--Kazanas Solution}

\author{Marina--Aura Dariescu\footnote{Email: marina@uaic.ro} and Ciprian Dariescu\footnote{Email: ciprian.dariescu@uaic.ro} \\
Faculty of Physics,
^^ ^^ Alexandru Ioan Cuza'' University of Ia\c{s}i \\
Bd. Carol I, no. 11, 700506 Ia\c{s}i, Romania}

\date{}
\maketitle

\begin{abstract}
We are considering the spacetime described by the metric proposed by Mannheim and Kazanas. The effective potential and the circular orbits are discussed. The rotational velocity derived from the geodesics equation agrees with the observed flat galactic rotation curves. 
Finally, solutions to the Gordon equation for massless bosons evolving in this spacetime are obtained in terms of Heun general functions.
\end{abstract}

\begin{flushleft}
{\it Keywords}: Mannheim--Kazanas Spacetime; Black Holes; Gordon equation; Heun functions.
\end{flushleft}

\begin{flushleft} 
{\it PACS:}
04.20.Jb Exact solutions;
02.40.Ky Riemannian geometries;
04.62.+v Quantum fields in curved spacetimes;
02.30. Gp Special functions.
\end{flushleft}

\newpage

\section{Introduction}

About thirty years ago, the Mannheim-Kazanas (MK) metric has been obtained as an exact exterior solution to conformal Weyl gravity
associated with a static, spherically symmetric gravitational source (Mannheim, 1989). The Schwarzschild term $2M/r$ is accompanied by a linear potential, $\gamma r$, and a quadratic contribution, of the form $- \lambda r^2$.
The linear term is generated
through the effect of cosmology on individual galaxies, while the quadratic one is induced by inhomogeneities in the cosmic background. These inhomogeneities are associated with distances larger than 1 Mpc.
Since the two contributions have opposed signs, the 
galaxies are able to support bound orbits. 

Even though the MK metric has not been obtained in Einstein's General Relativity, it has the advantage of solving the problem of the shape of the galactic rotation curves (Rubin, 1978), without postulating the dark matter existence.

As it is known, in real galaxies, at large values of $r$, the rotational velocity remains at nearly the same level or
is increasing.
Thus, in order to maintain the high
velocity, an additional contribution to the Newtonian one is required.

In this respect, the MK metric agrees with the empirical expression of the orbiting velocities of the visible matter around the center of galaxy 
\[
v^2 = \frac{a}{R} + b R \, ,
\]
where $R$ is the distance to the center of the galaxy, while the parameters $a$ and $b$ contain the number of stars in the galaxy.

In the last years, the MK  metric was tested on a broad sample of galaxies whose rotation curves extend
well beyond the galactic optical disks. The universal quantity $\gamma / \lambda$ gives a natural limit on the size of galaxies (Mannheim, 2011).

The present work is devoted to particles evolving on the Mannheim-Kazanas spacetime.
The effective potential and the conditions to have closed trajectories are discussed.
Working within a tetradic approach, the solutions to the Gordon equation are obtained, for different ranges of the radial variable.

The study of the motions of particles in this metric is very important since it provided a way to detect the presence of a global de Sitter-like
component and to found a specific value for its strength (Mannheim, 2011).

\section{The Mannheim-Kazanas metric and the Galaxy rotation curves}

Let us start with the original expression of the Mannheim-Kazanas (MK) vacuum solution (Mannheim, 1989)
\begin{equation}
g_{00} = 1-3 \beta \gamma - \frac{\beta (2-3 \beta \gamma)}{r} + \gamma r - \lambda r^2\, ,
\end{equation}
where $\gamma$ and $\lambda$ are universal constants, the same for all galaxies,
$\gamma = \gamma_0 = 3 \times 10^{-28} m^{-1}$ and $\lambda = 9.54 \times 10^{-50} m^{-2}$.

Also, $\beta =M$ can be associated to the mass
at the galaxy center, and one may work in the approximation $3 \gamma M \ll 1$.

The metric (1) can be written in the more transparent physical expression 
\begin{equation}
g_{00} = 1 -\frac{2M}{r} + \gamma r - \lambda r^2 
\end{equation}
and the spherically symmetric line element
\begin{equation}
ds^2= g_{11} (dr)^2 + r^2 \left[  (d \theta )^2 + \sin^2 \theta (d \varphi)^2 \right] - g_{00} (dt)^2 \; ,
\end{equation}
for $g_{11}= g_{00}^{-1}$, is given by
\begin{equation}
ds^2= \frac{(dr)^2}{g_{00}} + r^2 \left[  (d \theta )^2 + \sin^2 \theta (d \varphi)^2 \right] - g_{00} (dt)^2 = -d \tau^2 \; ,
\end{equation}
where $\tau$ is the proper time. 

Using the Lagrangian
\[
- L = \frac{1}{2} \left[ \frac{(\dot{r})^2}{g_{00}} + r^2 \dot{\theta}^2 + \sin^2 \theta \dot{\varphi}^2- g_{00} \dot{t}^2 \right]  ,
\]
one can derive the conserved energy and angular momentum
\begin{equation}
g_{00} \dot{t} = E \; , \;
r^2 \sin^2 \theta \dot {\varphi} = - \, K \, ,
\end{equation}
where {\it dot} means the derivatives with respect to $\tau$.

If the motion of the particle with zero angular momentum is on the equatorial plane, i.e. $\dot{\theta} = \dot{ \varphi} =0$, the relations (4) and (5) lead to the important constraint
\[
g_{00} \dot{t}^2 - \frac{\dot{r}^2}{g_{00}} =1 \, ,
\]
i.e.
\begin{equation}
\dot{r}^2 = E^2 - g_{00}  \, .
\end{equation}
In the above expression, one may identify the effective potential
\begin{equation}
V_{eff} = g_{00} = 1 - \frac{2M}{r} + \gamma r - \lambda r^2 \; ,
\end{equation}
which is represented in the figure 1. The potential goes to $- \infty$, for $r \to \pm \infty$ and has a singularity in $r=0$. 

\begin{figure}
  \centering
  \includegraphics[width=0.45\textwidth]{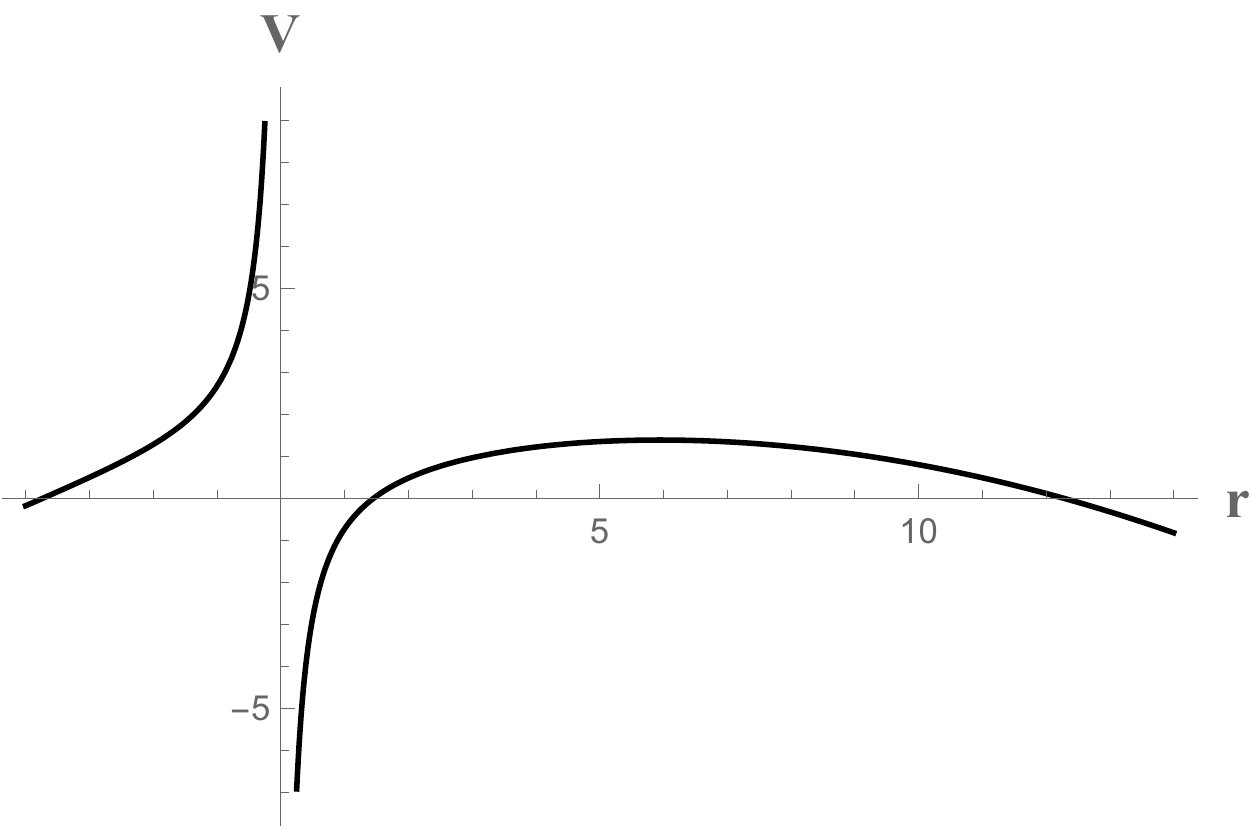} 
  \caption{The effective potential (7), for $\Delta >0$.} 
\end{figure}

The force acting on a particle, 
\[
F = - V_{eff}^{\prime} = - \frac{2M}{r^2} - \gamma + 2 \lambda r \, ,
\]
contains besides the attractive contributions, the repulsive term $\lambda r^2$ which becomes important at cosmological distances.

For the metric (2), the horizons are given by the equation $g_{00}=0$, namely they are the solutions of the cubic equation
\begin{equation}
- \lambda  r^3 + \gamma r^2 + r - 2M  =0 \; .
\end{equation}
This has three real roots if
the discriminant $\Delta = 18abcd-4b^3d+b^2c^2-4ac^3-27a^2d^2$ is a positive quantity (Shelbey, 1969),
i.e. 
\[
\Delta = -108 M^2 \lambda^2  + 4(1+9 \gamma M) \lambda + \gamma^2(1+8 \gamma M) > 0 \,  .
\]
To first order in $\gamma M \ll 1$ and $\lambda M^2 \ll1$, the three roots of (8) are given by the simple expressions: .
\begin{eqnarray}
r_1 \approx - \frac{1}{\gamma} \left[ 1 - \frac{\lambda}{\gamma^2} \right] ; \; \;
r_2 \approx 2 M \left[ 1 + 4 \lambda M^2 - 2 \gamma M \right] ; \; \; r_3 = \frac{\gamma}{\lambda}  + \frac{1}{\gamma}  \; .
\end{eqnarray}
Thus, there are two positive roots of the equation (8), corresponding to the physical horizons.
The first one is the black hole horizon, $r_h = r_2$, situated in the Schwarzschild region, while the second one corresponds to the cosmological horizon, $r_{\lambda} = r_3$.

The existence of a circular orbit of radius $R_c$ imposes the conditions
$\dot{r}=0$ and $\dot{r}^{\prime} =0$, at $r= R_c = const.$ The corresponding equation  
\[
V_{eff}^{\prime} ( r= R_c ) =0 \,  ,
\]
i.e.
\begin{equation}
-2 \lambda R_c^3 + \gamma R_c^2 + 2M =0 
\end{equation}
has a negative discriminant $\Delta = -8M (\gamma^3 + 54 \lambda^2 M)$ and therefore we have one real solution
\[
R_c = \frac{1}{6 \lambda} \left[ \gamma + \frac{\gamma^2}{K^{1/3}} + K^{1/3} \right] \; , \; \;
K = \gamma^3 + 108 \lambda^2 M - \left[ 216 \lambda^2 M \left( \gamma^3 + 54 \lambda^2 M \right) \right]^{1/2} .
\]
To first order in $M \lambda^2 / \gamma^3$, the circular orbit's radius is
\begin{equation}
R_c \approx \frac{\gamma}{2 \lambda} + \frac{4M \lambda}{\gamma^2} \, ,
\end{equation}
and it depends, besides the universal constants $\gamma$ and $\lambda$, on the mass $M$.

Thus, a particle with the energy
\[
E^2  = V_{eff} (r= R_c )  = V_{max} 
\]
will follow a circular orbit of radius $R_c$, with $R_c$ in between the two horizons.
This circular orbit is unstable because the potential is concave down.

For particles with the energy $E^2 < V_{max}$, there may be two turning points (see figure 2). If one imposes $\dot{r}=0$ in the relation (6), he obtains the equation
\[
E^2 = 1 - \frac{2M}{r} + \gamma r - \lambda r^2 \, .
\]
\begin{figure}
  \centering
  \includegraphics[width=0.45\textwidth]{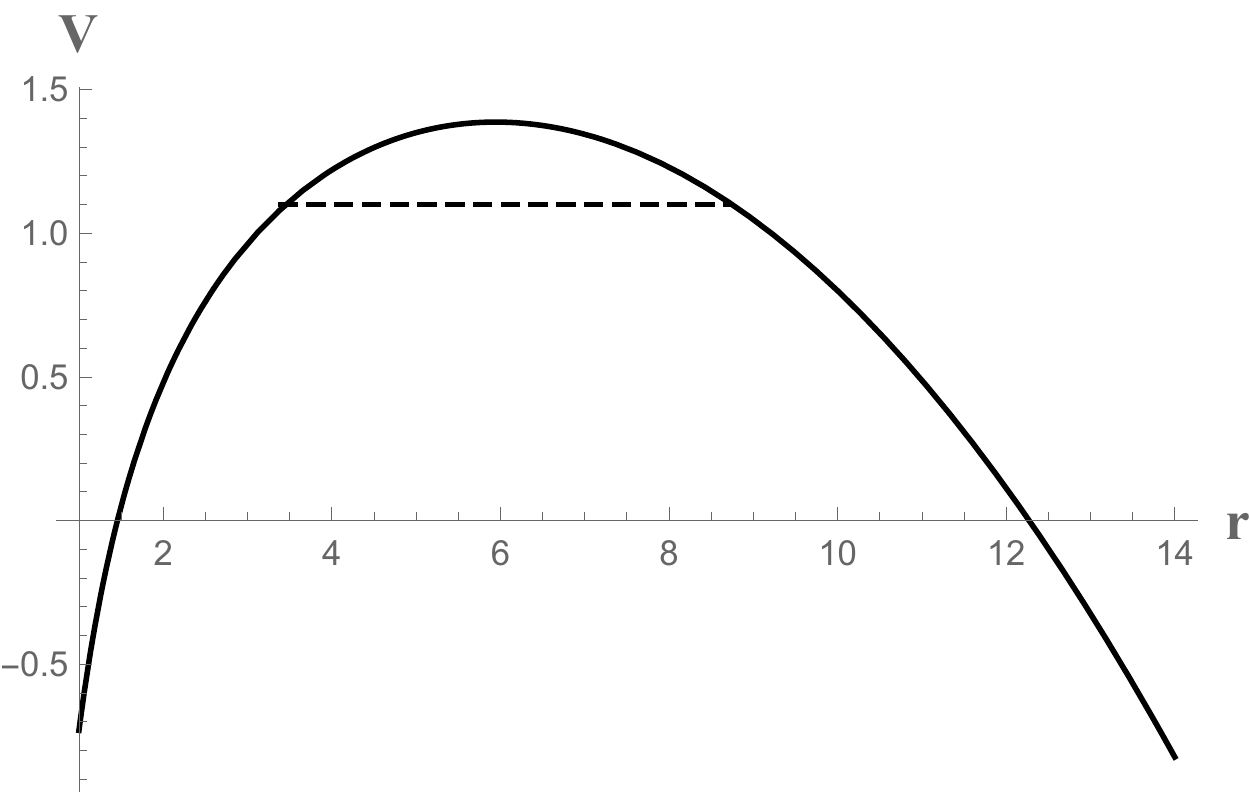} 
  \caption{The particle with the energy $E^2 < V_{max}$ is represented by the dashed line. The two turning point are given by the intersections with the effective potential.} 
\end{figure}
The corresponding cubic equation
\begin{equation}
\lambda r^3 - \gamma r^2 + \varepsilon r + 2M =0 \, ,
\end{equation}
with $\varepsilon = E^2 -1$, has three real solution if $\Delta >0$. To first order in $\varepsilon$, this means the energy range
\[
0 \leq \varepsilon < \frac{2 \gamma^3 - 27 \lambda^2 M}{9 \gamma \lambda} \; ,
\]
for $M< 2 \gamma^3 /(27 \lambda^2) \approx 2 \times 10^{14} m \approx 10^{11} M_S$.

The two positive roots of (12), denoted by $r_1$ and $r_2$, are situated in between the two horizons. In the figure 2, these are given by the intersections of the dashed line with the potential. When the energy is increasing, the interval between $r_1$ and $r_2$ becomes smaller and the elliptic trajectory turns into a circle for $E^2 = V_{max}$.
The relations between the important values of $r$ discussed up to now are:
$r_h < r_1 < R_c < r_2 < r_{\lambda}$.

Let us mention that a particle coming from large $r$ values, on her way to the horizon $r_h$, is encountering a Minkowskian region where the terms $M/r$ and $\lambda r^2$ are compensated by $\gamma r$ and therefore $g_{00} \approx 1$. The positive real solutions of the equation $g_{00}=1$, i.e.
\[
- \lambda r^3 + \gamma r^2 - 2M =0  \; ,
\]
whose $\Delta$ is positive for $2 \gamma^3 > 27 \lambda^2 M$,
can be approximated to
\begin{equation}
R_{1} \approx \frac{M \lambda}{\gamma^2} + \sqrt{\frac{2M}{\gamma}}  \, , \; \;
R_{2} \approx \frac{\gamma}{\lambda} - \frac{2M \lambda}{\gamma^2} \, .
\end{equation}
One can check that $R_1 \ll R_2$ and both $R_1$ and $R_2$ are in between the two horizons in (9). Also, for large $r$ values, where the $M/r$ contribution can be neglected, one has only one universal Minkowskian region, at $R = \gamma / \lambda$, which gives the size of galaxies.

Finally, let us discuss the intriguing subject of
orbiting velocities of matter around the galaxy center.

We are using a free of coordinates approach and introduce, for the line element (3), the pseudo-orthonormal frame ${E_a}_{(a=\overline{1,4})}$
\begin{equation}
E_1=  \sqrt{g_{00}} \, \partial_r \; , \; \;
E_2= \frac{1}{r}  \, \partial_{\theta} \; , 
\; \;
E_3=\frac{1}{r \sin \theta}\, \partial_{\varphi} \; , \; \;
E_4= \frac{1}{\sqrt{g_{00}}} \, \partial_t \; ,
\end{equation}
whose corresponding dual base is 
\begin{equation}
\omega^1=  \frac{dr}{\sqrt{g_{00}}} \; , \; \;
\omega^2=  r  \,d\theta \; , 
\; \;
\omega^3= r \sin \theta \, d\varphi \; , \; \;
\omega^4= \sqrt{g_{00}} \,dt \; ,
\end{equation}
so that $ds^2 = \eta_{ab} \omega^a \omega^b$, with $\eta_{ab} = {\rm diag} [ 1 , 1 , 1 , -1]$.

Using the first Cartan's equation,
\begin{eqnarray}
& &
d\omega^a=\Gamma^a_{.[bc]}\,\omega^b\wedge \omega^c \;  ,
\end{eqnarray}
with $1 \leq b<c \leq4$ and $\Gamma^a_{.[bc]} = \Gamma^a_{.bc} - \Gamma^a_{. cb}$,
we obtain the connection coefficients
\begin{eqnarray}
& &
\Gamma_{122} =  \Gamma_{133} = - \frac{ \sqrt{g_{00}}}{r} = - \Gamma_{212}  = - \Gamma_{313} 
\nonumber \\*
& &
\Gamma_{233} = - \Gamma_{323} = - \, \frac{\cot \theta}{r}  \; ,
\nonumber \\*
& &
\Gamma_{144} = - \Gamma_{414} = \frac{g_{00}^{\prime}}{2 \sqrt{g_{00}}} \; .
\end{eqnarray}

The equations for the timelike geodesics
\[
\frac{du^a}{d \tau} + \Gamma^a_{\; bc} u^b u^c = 0 \;  ,
\]
where the four velocity components, 
\[
u^a = \frac{\omega^a}{d \tau} \, ,
\]
are given by:
\[
u^1 = \frac{\dot{r}}{\sqrt{g_{00}}} \; , \; \; u^2 = r \dot{\theta} \; , \; \; u^3 = r \sin \theta \dot{\varphi} \; , \; \;
u^4 = \sqrt{g_{00}} \dot{t} \; ,
\]
have the explicit form
\begin{eqnarray}
& &
\ddot{r} - \frac{g_{00}^{\prime}}{2 g_{00}} \dot{r}^2 - r g_{00} \left[ \dot{\theta}^2 + \sin^2 \theta \dot{\varphi}^2 \right] + \frac{g_{00} g_{00}^{\prime}}{2} \, \dot{t}^2 = 0  \; ,
\nonumber \\*
& & r \ddot{\theta} + 2 \dot{r} \dot { \theta} - r \sin \theta \cos \theta \dot{\varphi}^2 = 0 \; , \nonumber \\*
& & r \sin \theta \ddot{\varphi} + 2 \sin \theta \dot{r} \dot{ \varphi} + 2 r \cos \theta \dot{\theta} \dot{ \varphi} = 0 \, ,
\nonumber \\*
& & \ddot{t} + \frac{g_{00}^{\prime}}{g_{00}} \, \dot{r} \dot{t} = 0  \;  .
\end{eqnarray}
For the circular orbit in the equatorial plane, i.e. 
$\dot{r}=0$, $\theta = \pi/2$, $\dot{\theta} =0$, the first relation in (18) becomes
\[
\frac{d \varphi}{dt} = \sqrt{\frac{g_{00}^{\prime}}{2r}} \,  ,
\]
leading to the velocity
\begin{equation}
v = r \frac{d \varphi}{dt} = \sqrt{\frac{r g_{00}^{\prime}}{2}} \,  .
\end{equation}
Obviously, in the Schwarzschild case, one obtains the well known Newtonian expression
\[
v = \sqrt{\frac{M}{r}} \, ,
\]
while for the metric (2), the velocity is given by
\begin{equation}
v = \sqrt{\frac{1}{2} \left[ \frac{2M}{r} + \gamma r - 2 \lambda r^2 \right]} .
\end{equation}
For $2 \lambda r \ll \gamma$, when only the first two terms in the parenthesis are taken into account, one may notice that the velocity is significantly increased, compared to the Schwarzschild  expression.

The relation $v^2 \geq 0$ is satisfied by $r \leq R_c$, where $R_c$ is the circular radius (11). 

\section{The Gordon equation and its Heun solutions}

Working in a free of coordinates approach (Dariescu, 2017), based on Cartan's equations, the
Einstein tensor components in the tetradic frame (14) are: 
\begin{eqnarray}
& &
G_{11} = \frac{2 \gamma}{r} - 3 \lambda \; , \nonumber \\*
& &
G_{22} = G_{33} = \frac{\gamma}{r} - 3 \lambda \;   , \nonumber  \\*
& & G_{44} = 3 \lambda - \frac{2\gamma}{r} \; .
\end{eqnarray}
One may notice that, from the point of view of Einstein's General Relativity, the proper energy density $\rho = T_{44} =G_{44}$ is negative for $r< ( 2 \gamma)/(3 \lambda)$, leading to the conclusion that, at values of $r$ smaller than $2 \gamma / ( 3 \lambda)$, the astrophysical object described by the MK metric is surrounded by exotic matter.

The scalar curvature,
\[
R = 12 \lambda - \, \frac{6\gamma}{r} \, ,
\]
has a singularity in $r=0$ and is positive for $r> \gamma/(2 \lambda)$.

In the general expression of the Klein-Gordon equation (Dariescu, 2017)
\begin{eqnarray}
& &
\frac{1}{r^2} \frac{\partial \;}{\partial r} \left[ r^2 g_{00} \frac{\partial \Phi}{\partial r} \right]
+ \frac{1}{r^2 \sin \theta} \frac{\partial \;}{\partial \theta} \left[ \sin \theta \frac{\partial \Phi}{\partial \theta} \right]
+ \frac{1}{r^2 \sin^2 \theta} \frac{\partial^2 \Phi}{\partial \varphi^2}
\nonumber \\*
& & - \; 
\frac{1}{g_{00}} \frac{\partial^2 \Phi}{\partial t^2} - \mu^2 \Phi \, = 0 \; ,
\end{eqnarray}
where $g_{00}$ given in (2) is depending only on $r$, one can perform the variables separation
\begin{equation}
\Phi = F(r) Y_{\ell}^m ( \theta , \varphi ) \, e^{-i \omega t} \, ,
\end{equation}
where $Y_{\ell}^m$ are the spherical functions. The corresponding radial equation,
\begin{eqnarray}
\frac{1}{r^2} \frac{d \;}{dr} \left[ r^2 g_{00} \frac{dF}{dr} \right]
+ \left[ \frac{\omega^2}{g_{00}} - \frac{\ell ( \ell+1) }{r^2} - \mu^2 \right] F = 0 \, ,
\end{eqnarray}
can be analytically solved only in particular cases.

Let us consider the very large distances, typical to clusters and superclusters, for which the metric function can be approximated to
\begin{equation}
g_{00} \approx 1 +\gamma r - \lambda r^2  = - \lambda \left[ r + \frac{b-\gamma}{2\lambda} \right] \left[ r - \frac{b+ \gamma}{2 \lambda} \right] ,
\end{equation}
where $b = \sqrt{\gamma^2 + 4 \lambda}$.
There is an unique horizon, which is the positive root of the equation $g_{00}=0$, i.e.
\begin{equation}
r_h = \frac{b+\gamma}{2 \lambda}  \; ,
\end{equation}
which can be understood as the boundary of this spacetime.
Using the numerical values of $\gamma = \gamma_0$ and $\lambda$, we get $r_h \approx 100 Mpc$.

The radial function, solution to (24) with $\mu=0$, is expressed in terms of Heun general functions (Ronveaux, 1995; Slavyanov, 2000), as
\begin{eqnarray}
F(r)  & = & \left[ r -  \frac{b + \gamma}{2 \lambda} \right]^{-4i \omega \lambda/b}  \left[ r +  \frac{b - \gamma}{2 \lambda} \right]^{4i \omega \lambda/b}  r^{- \frac{1 \pm \sqrt{1+16 \lambda \ell ( \ell +1)}}{2}}  \nonumber \\* & \times & 
HeunG \left[ {\rm a}  \, , q , \, \alpha , \, \beta , \, \gamma , \, \delta ,  - ( \lambda r_h ) r  \right] .
\end{eqnarray}

As it is known from the theory (Slavyanov, 2000), the Heun general function of variable $x$ has four singular points and one may check that $r= r_h$ is one of them.

As we turn to regions well inside a galactic halo boundary, the $\lambda$ contribution can be neglected and the term $2M/r$ becomes important.
The metric (2) can be approximated to
\begin{equation}
g_{00} = 1- \frac{2M}{r} + \gamma r = \frac{\gamma}{r} \left[ r +  \frac{a+1}{2\gamma} \right]  \left[ r -  \frac{a-1}{2\gamma} \right] ,
\end{equation}
where
\[
a = \sqrt{1+8 \gamma M}   \, .
\]
This case has been investigated, in detail, in (Dariescu, in press).
Similarly to the previous case, there is one horizon, close to the Schwarzschild radius,
\begin{equation}
r_h = \frac{a-1}{2 \gamma} \approx 2M ( 1-2 \gamma M) \; .
\end{equation}

In the massless case, the radial function is expressed in terms of Heun general functions (Ronveaux, 1995; Slavyanov, 2000), as 
\begin{eqnarray}
F(r)  =  \left[ r -  \frac{a-1}{2 \gamma} \right]^{\frac{ i \omega (a-1)}{2 \gamma a}}\left[ r +  \frac{a+1}{2 \gamma} \right]^{- \, \frac{ i \omega (a+1)}{2 \gamma a}}
HeunG \left[ {\rm a} \, , q , \, \alpha , \, \beta , \, \gamma , \, \delta , 1 + \frac{2 \gamma r}{a+1} \right] .
\end{eqnarray}

The function (30) is much more complex than the one obtained for the Schwarzschild metric 
\[
g_{00} = 1 - \frac{2M}{r} \, ,
\]
where the amplitude function expressed in terms of the Heun confluent functions as
\begin{equation}
F_S = e^{i \omega r} ( r-2M)^{\mp 2i \omega M} HeunC \left[ - 4i \omega M , \mp 4 i \omega M, 0 , - 8 \omega^2 M^2 , 8 \omega^2 M^2 - \ell (\ell+1), 1 - \frac{r}{2M} \right] .
\end{equation}

Compared to the Heun general functions which have four singular points, the Heun confluent functions have two regular and one irregular singularities and these can be obtained from the Heun general functions by a confluence process, when two of the singularities coalesce.
The Heun confluent function can be computed as a power series expansion around the origin $z=0$, i.e. $r=2M$, and the series converges for $|z| <1$,
 (Ronveaux, 1995; Slavyanov, 2000).

The horizon defined in (29) is closer to the $r=0$ singularity compared to the Schwarszchild horizon, $r_S =2M$.
On her way to the horizon $r_h$, the particle is crossing the Minkowskian region $r=R = \sqrt{2M/\gamma}$, where $g_{00} \approx 1$.

One may notice that, when $r \to 0$, the amplitude function (30) goes to infinity since the Heun general function, of variable $x$, has a singularity in $x=1$.
However, we should not worry about that because the metric (2) is valid only on the external region of an astrophysical object of radius $R$.

\section{Conclusions} 

In this work, we have considered a spacetime described by the metric proposed by Mannheim and Kazanas (MK metric) (Mannheim, 1989), which contains two parameters, $\gamma$ and $\lambda$. We have worked with the universal values: $\gamma = \gamma_0 = 3 \times 10^{-28} m^{-1}$ and $\lambda = 9.54 \times 10^{-50} m^{-2}$, so that $\gamma / \lambda \approx 100 kpc$ (Mannheim, 2011). Obviously, this value of $\gamma$ is independent of the galactic mass. 

At a first site, the MK metric can be seen as an extension of the Schwarzschild solution. However, one can notice significant differences between these two.
Thus, for the metric (2), the horizons are given by the roots of the cubic equation (8) and have the approximate expressions given in (9). One can identify a Minkowskian region, in between the two horizons, where $g_{00} \approx 1$, which has no analogue to the Schwarzschild case. The constant potential is leading to approximately constant rotational velocities which agree with the observed flat galactic rotation curves (Rubin, 1978).

For a typical galaxy with $M = 10^{11} M_s$, so that
\[
M = \frac{10^{11} M_S G}{c^2} \approx 1.5 \times 10^{14} m \, ,
\]
the terms $2M/r$ and $\gamma r$ are almost equal for $r_* = \sqrt{2M / \gamma } \approx 30 kpc$, which corresponds to the Minkowskian region, once we neglect the $\lambda r^2$ contribution.

Also, a particle moving in the potential (7), with suitable energy values, can follow an elliptic or a circular orbit. The radius (11) has been calculated in the approximation $M \lambda^2 / \gamma^3 \approx 3M \times 10^{-16} \ll 1$. The universal quantity $\gamma / \lambda$ has been seen as a natural geometric limit on the size of galaxies, in the sense that beyond this distance there could no longer
be any bound galactic orbits (Mannheim, 2011). In addition, a very important point has been made in (Nandi, 2012), with respect to the actual size of the galaxies, which must be related to the radius of the last circular stable orbit. 

The velocity expression (20), derived from the geodesics equation, agrees with the one obtained by O'Brien and Mannheim in the framework of conformal gravity theory (O'Brien, 2012).
As it is known, the Newtonian velocity falls below data as the orbital radius increases. The presence of the linear term is leading to an excess in the observed velocities.
With no free parameters
other than the galactic mass, O'Brien and Mannheim proved that their result can fit an impressive number of galaxies.

However, for fitting the observed orbital velocities for more than 100 galaxies, $\gamma$ has been taken as $\gamma = \gamma_0 + \gamma_G$, where $\gamma_G = N \gamma^*$, with $N = M/M_S$ and $\gamma^* = 5.42 \times 10^{-39} m^{-1}$ (Mannheim, 2012).
Thus, besides the global cosmological linear term, there is a second one, which is depending on the matter within the galaxy.
Such an analysis is beyond the purpose of our paper and for a detailed discussion on this subject, we recommend (Nesbet, 2018).

In the final part of the paper, we have worked out the Gordon equation for massless bosons evolving in the MK spacetime. The radial equation is considered for different regions of the variable $r$, starting with distance scales between $1 Mpc$ and
$100 Mpc$, where $g_{00}$ can be taken as in (25), down to regions where the $2M/r$ term has a significant contribution and the term $\lambda r^2$ can be neglected.

The whole analysis is much more involved compared to the Schwarzschild exterior metric, the solutions to the Gordon equation being given by the Heun general functions, (Slavyanov, 2000).

\end{document}